# Design and Realization of the LST Main Structure for the Cherenkov Telescope Array

T. Schweizer*[a], J. Eder+[a], H. Wetteskind[a], C. Jablonski[a], R. Stadler[a],
R. Mirzoyan[a], D. Paneque[a], T. Masahiro[a],
A. Fiasson[b], N. Geffroy[b], G. Deleglise[b], O. Blanch[c], J. Mundet[c], R. Garcia[c],
A. J. Penuela, Lopez[d]
[a]Max Planck Institut für Physik, Boltzmannstr. 8, 85748 Garching, Germany; [b]Laboratoire d'Annecy de Physique des Particules (LAPP), 9 Chemin de Bellevue, 74940 Annecy-le-Vieux, France; [c]Institut de Física d'Altes Energies (IFAE), Edifici Cn, Campus UAB 08193 Bellaterra (Barcelona), Spain; [d]Universidad de Jaén, Campus Las Lagunillas, Ctra. Torrequebradilla s/n, 23071 Jaén, Spanien

## ABSTRACT

The 23 m diameter Large Size Cherenkov Telescope (LST) for CTA, located at 2250 m a.s.l. on the Canary Island of La Palma, is the next-generation Cherenkov telescope following MAGIC, H.E.S.S., and VERITAS. To enable rapid repositioning (180° in 18 s) for gamma-ray burst observations, the mechanical structure was designed to be ultra-lightweight (110 tons). The space-frame structure consists of slender struts made of carbon fibre, aluminium (dish and camera mast), and steel. The telescope is designed to withstand extreme environmental conditions at the ORM observatory on La Palma, including wind speeds up to 200 km/h, uplift forces, and ice loads of up to 30 tons. We present the structural design developed to meet these functional and environmental requirements. MPP Munich is responsible for the telescope's mechanical structure together with partner institutes in France (LAPP) and Spain (IFAE). The prototype, LST-1, has been operational since 2019, and three additional LSTs are curre[1]ntly under construction.

**Keywords:** Cherenkov Telescope, Gamma Ray Astrophysics, LST, ORM, La Palma, CTA

## 1. INTRODUCTION

The 23-meter-diameter Large-Sized Telescope (LST) of the Cherenkov Telescope Array (CTA) was developed during an approximately eight-year design and engineering phase between 2008 and 2016. The first telescope, LST-1, was successfully constructed and commissioned at the Observatorio del Roque de los Muchachos (ORM) on the Canary Island of La Palma at an altitude of 2,200 m above sea level. Since 2019, it has been in continuous operation and, as of 2026, regularly acquires scientific data alongside the MAGIC telescopes. The outstanding performance of the LST design has led to its replication, and three additional telescopes (LST-2 to LST-4) are currently being installed and commissioned. Together, the four telescopes will form the world's most sensitive Cherenkov telescope array in the low-energy gamma-ray regime, achieving an unprecedented energy threshold of approximately 20 GeV. Figure 1 shows the LST at the ORM site on La Palma.

The LST is the largest instrument of CTA and is specifically optimized for the detection of very-high-energy gamma rays at the lowest energies accessible from the ground. Achieving this scientific objective requires a large 23 m diameter reflector while maintaining a highly dynamic structure capable of rapid repositioning.

*tschweizlmpp.mpg.de; phone +49 89 32354 227; http://mpp.mpg.de,
+josef.eder@t-online.de, www.e-der.de

Observations of transient phenomena, such as gamma-ray bursts, require the telescope to repoint to any position in the sky within 20 s. At the same time, the structure must maintain the optical precision required for scientific observations under varying environmental conditions and over an operational lifetime exceeding 30 years. These requirements are particularly challenging because the telescope is installed at a high-altitude observatory where extreme wind loads, ice accretion, and seismic actions must be considered.

Meeting these demanding performance objectives required the development of an ultra-lightweight structural concept based on a modular space-frame architecture. The design builds upon the experience gained with the 17 m MAGIC telescopes, which demonstrated the advantages of lightweight truss structures for rapidly slewing Cherenkov telescopes. For the significantly larger LST, the structural concept was further optimized with respect to stiffness, mass, manufacturability, and operational reliability.

The mechanical structure of the LST was developed by the Max Planck Institute for Physics (MPP) in collaboration with MERO-TSK, a company with extensive expertise in large-span space-frame structures. The resulting telescope combines low structural mass, high dynamic performance, and robustness against extreme environmental loading. This paper presents the principal design requirements, the adopted structural solutions, and the engineering challenges encountered during the development of the LST mechanical structure.

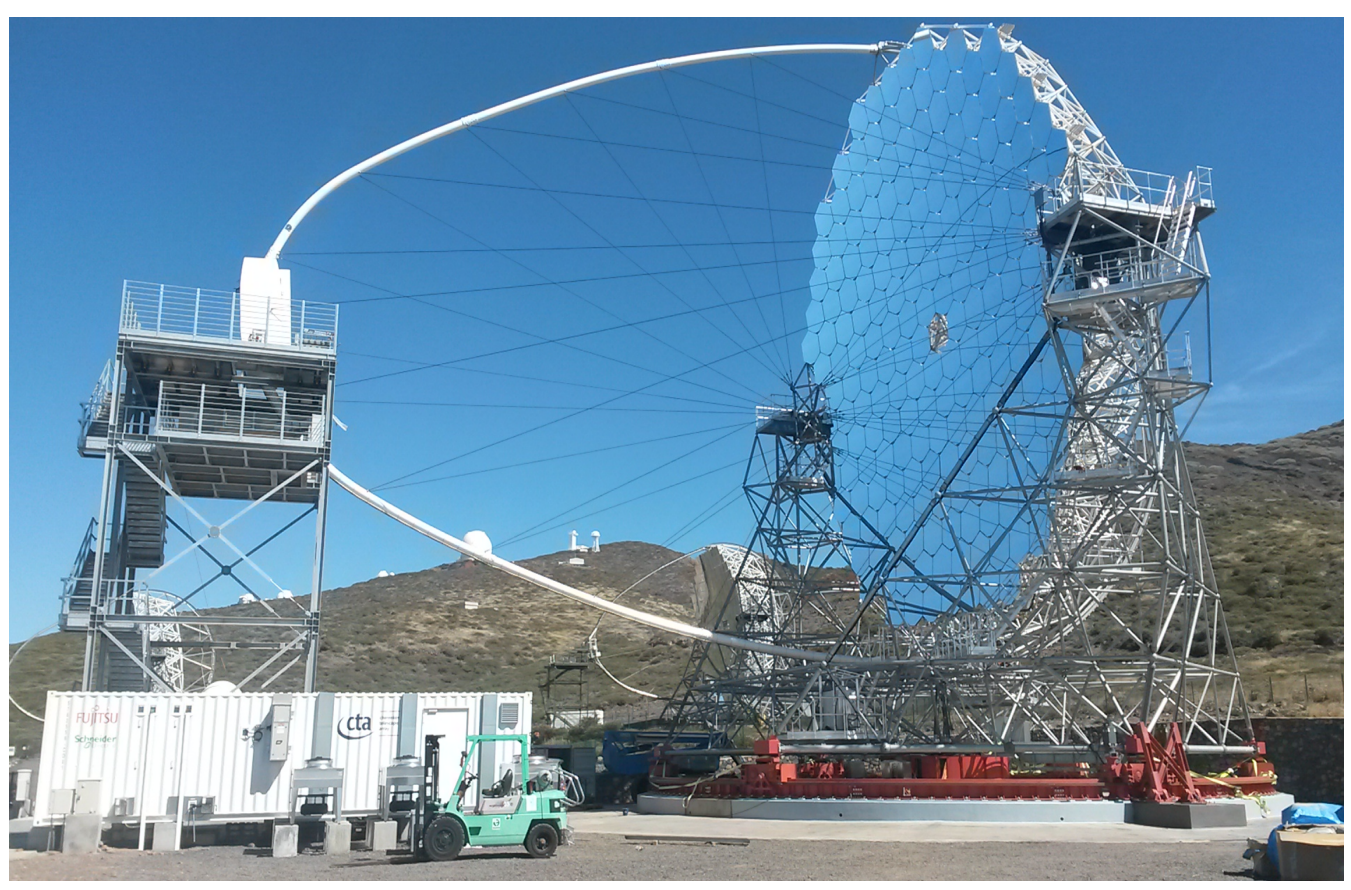

Figure 1. The first 23m diameter LST at the Roque de los Muchachos on the Canary Island of La Palma
(credits: LST collaboration)

## 2. DESIGN REQUIREMENTS AND CHALLENGES

Cherenkov telescopes differ from conventional astronomical telescopes because they observe the short-lived Cherenkov light produced by rapidly moving particle showers in the atmosphere. Large instruments such as H.E.S.S., MAGIC, and others are typically based on lightweight truss structures.

The principal design requirements for Cherenkov telescopes are:

- Rapid repositioning to acquire transient targets.
- High availability for scientific observations over a service life of at least 30 years.
- Low recurring cost to enable the construction of large telescope arrays.
- A segmented primary reflector equipped with active mirror alignment for each mirror segment.

A major challenge for the structural design is the stringent wind-load specification, including gust speeds of up to 200 km/h.

These requirements were already applicable to the MAGIC telescopes, which feature 17 m diameter primary mirrors and have been in operation since 2004. During their development, MPP collaborated with a company specializing in large-span truss structures for roofs, industrial buildings, and other complex civil engineering applications. The company, MERO-TSK (Würzburg, Germany), employed a modular system of standardized struts that enabled efficient fabrication and rapid assembly. Although primarily focused on civil structures, including airport and exhibition hall roofs, MERO-TSK agreed to support the development of the primary structure for the MAGIC telescopes and subsequently manufactured, delivered, and assembled the two telescopes at the Roque de los Muchachos Observatory on La Palma.

Building on this successful collaboration, MPP adopted the same design concept for the 23 m diameter LST reflector while incorporating improvements based on operational experience gained with MAGIC and the additional requirements imposed by the larger telescope size.

## 3. TELESCOPE DESIGN

The LST employs a conventional altitude–azimuth telescope configuration. The structural design and fabrication were carried out by MERO-TSK using their proprietary modular truss system. The structural components are certified for civil engineering applications and are subject to rigorous quality-control procedures.

The entire telescope is supported by the azimuth rail and drive system. The azimuth drive consists of four passive and two active bogies, each equipped with four wheels. Both the rail and bogies were specifically designed for the lightweight telescope structure and include anti-uplift provisions to resist extreme wind loads, as the telescope's self-weight alone is insufficient to counteract the maximum wind-induced uplift forces.

The 23 m diameter reflector comprises 198 hexagonal mirror segments manufactured as sandwich structures with cold-slumped glass face sheets and aluminium honeycomb cores. Each mirror segment is equipped with an active mirror control (AMC) system that provides precise alignment. The mirror panels are mounted to the dish structure through the AMC actuators.

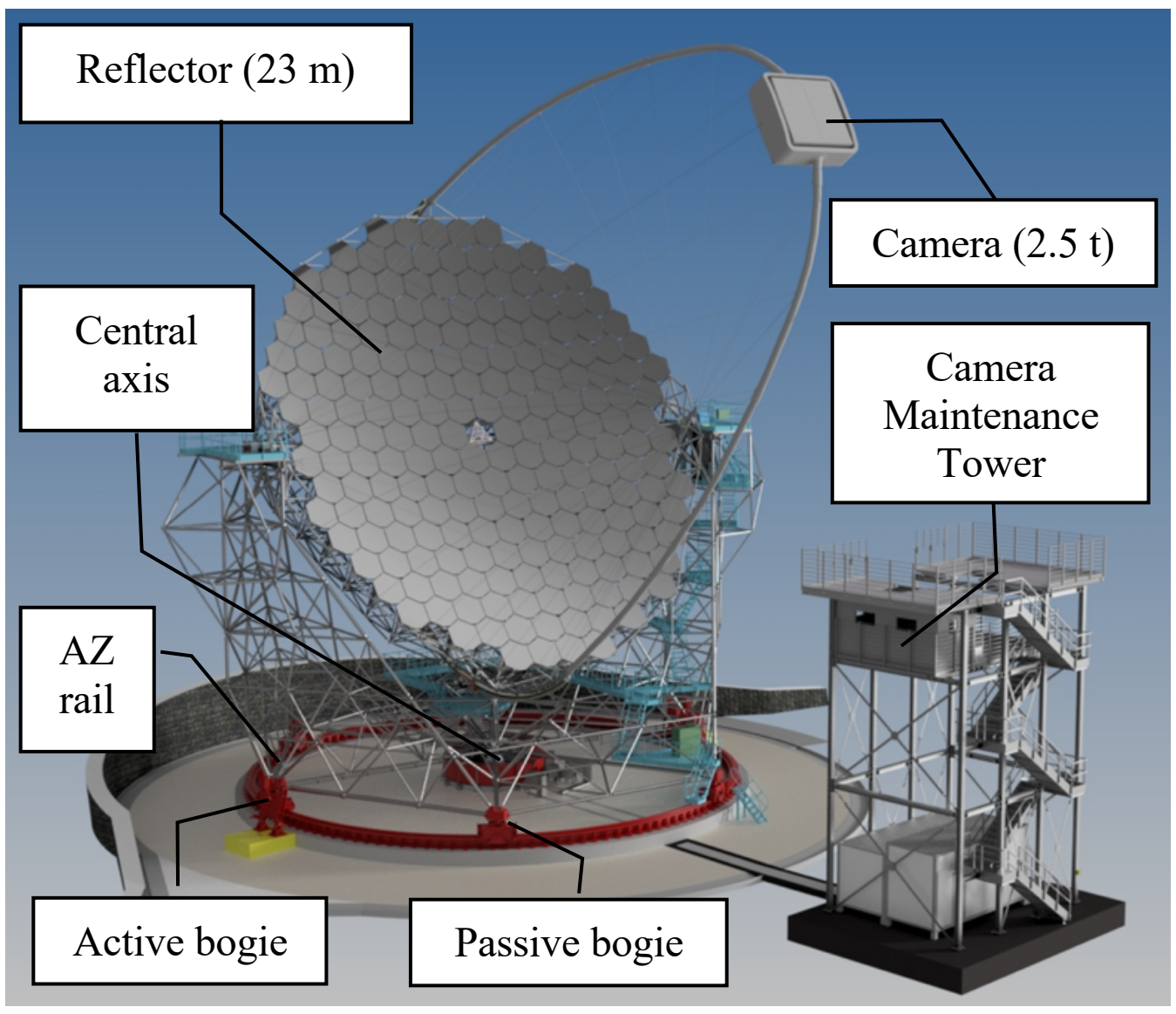

Figure 2. Mechanical FEM model of the LST

Passive bogie

Table 1: Main characteristics and design parameters of the LST

| The main characteristics and design requirements | |
|---|---|
| Envelope height in Zenit position | 46 m |
| Envelope height in parking position | 31 m |
| Envelope width | 32 m |
| Reflector diameter | 23 m |
| Reflector surface (net) | 386 $m^2$ |
| Focal length | 28 m |
| F/D | 1.2 |
| Camera FoV | 4.5° |
| Azimuth angle range | ±270 deg |
| Elevation angle range | -70 deg to +90 deg |
| Total rotating mass | 110 tons |
| Azimuthal inertia | 12000 tons*$m^2$ |
| Elevation inertia | 5600 tons*$m^2$ |
| Elevation drive max angular velocity | 72.6 µrad/s |
| Azimuth drive max angular velocity | 8.4 mrad/s |
| Max acceleration/emergency stop | 0.1 rad/$sec^2$ |
| Rotation time for 180 deg | <18 sec |

The 2.5-ton camera is positioned 28 m from the reflector vertex, which coincides with the elevation axis. It is supported by a camera frame constructed from carbon-fibre-reinforced polymer (CFRP) and stabilized by a system of 26 CFRP tension members. The complete elevation assembly, consisting of the reflector dish, camera support structure, and camera, is counterbalanced by the rear arch of the dish to enable smooth and rapid motion. The elevation drive is connected to the rear arch through a chain-drive mechanism. Table 1 lists the key design parameters and telescope characteristics of the LST.

The development and production of the entire LST is shared between several institutes. The responsibilities for the individual sub-system are listed in table 2.

Table 2: Institute responsibilities for the LST prototype mechanics:

| Institute responsibilities for the mechanical structure of the LST prototype | |
|---|---|
| Main structure (lower structure, dish, central pin): | MPP, Germany |
| Access system (stairs. Platforms, gangways): | MPP, Germany |
| Azimut and elevation cable chains: | MPP, Germany |
| Camera support structure (arch, camera frame): | LAPP, France |
| Camera support structure tether (ropes and fittings): | INFN, Italy |
| Azimuth rail: | MPP, Germany |
| Azimuth drive: | IFAE, Spain |
| Elevation drive: | MPP, Germany |
| Camera access tower: | MPP, Germany + UCAEN, Spain |
| Foundation: | IAC |

The participating institutes independently designed and manufactured their respective subsystems, outsourcing specific activities where appropriate. MPP was responsible for the overall mechanical system design and finite-element analyses, defining the subsystem interfaces as well as the corresponding design load cases and requirements.

## 3. MAIN STRUCTURE DESIGN

The main structure of the LST, which is the subject of this paper, is based on the MERO-TSK space-frame system and consists of two primary assemblies: the azimuth (AZ) structure, forming the lower support structure, and the elevation (EL) structure, which supports the reflector dish. Together, these assemblies constitute the telescope mount. The third major structural element, the Camera Support Structure (CSS), is not discussed in this paper, as it employs a different and equally interesting structural technology.

Figure 2 shows the assembled main structure. The lower AZ structure is erected directly on the azimuth drive system by assembling the truss strut-by-strut, starting with the spoke-wheel-type base layer mounted on the azimuth bogies. The EL structure is assembled separately at ground level and subsequently lifted by crane onto the elevation bearings located on the AZ structure platforms.

The figure also shows the CSS dummy equipped with a ballast mass that simulates the weight and center of gravity of the camera. This temporary assembly is required to balance the elevation structure and enable safe rotation of the EL assembly during integration and commissioning before installation of the camera. Without the camera or equivalent ballast, the elevation assembly would be subject to a significant mass imbalance. In addition, the reflector mirror panels can be seen already installed on the dish structure. During construction, these mirrors were protected by tarpaulin covers to prevent damage and contamination.

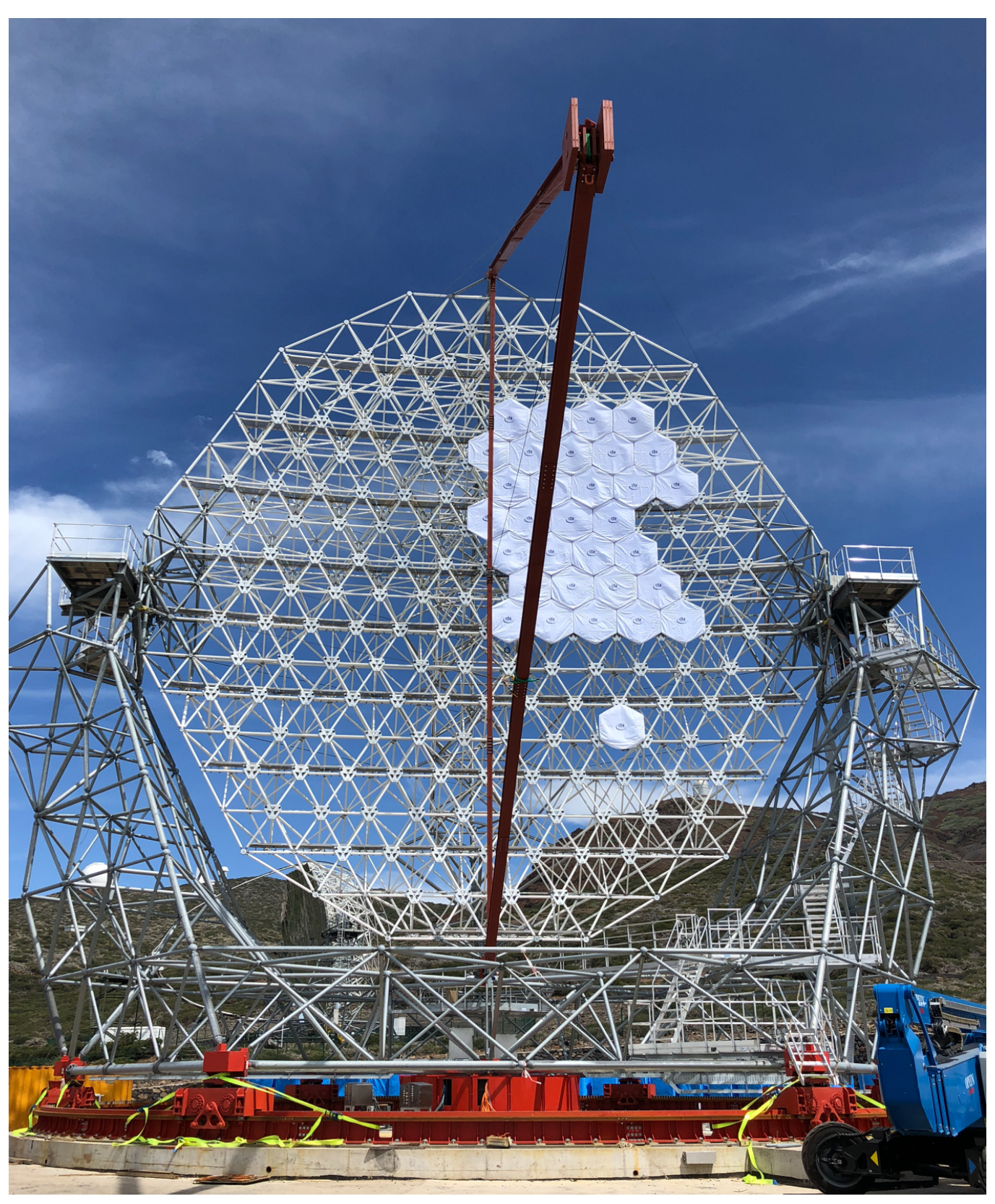

Figure 3. Main structure and drives of the LST with CSS dummy and some mirrors integrated

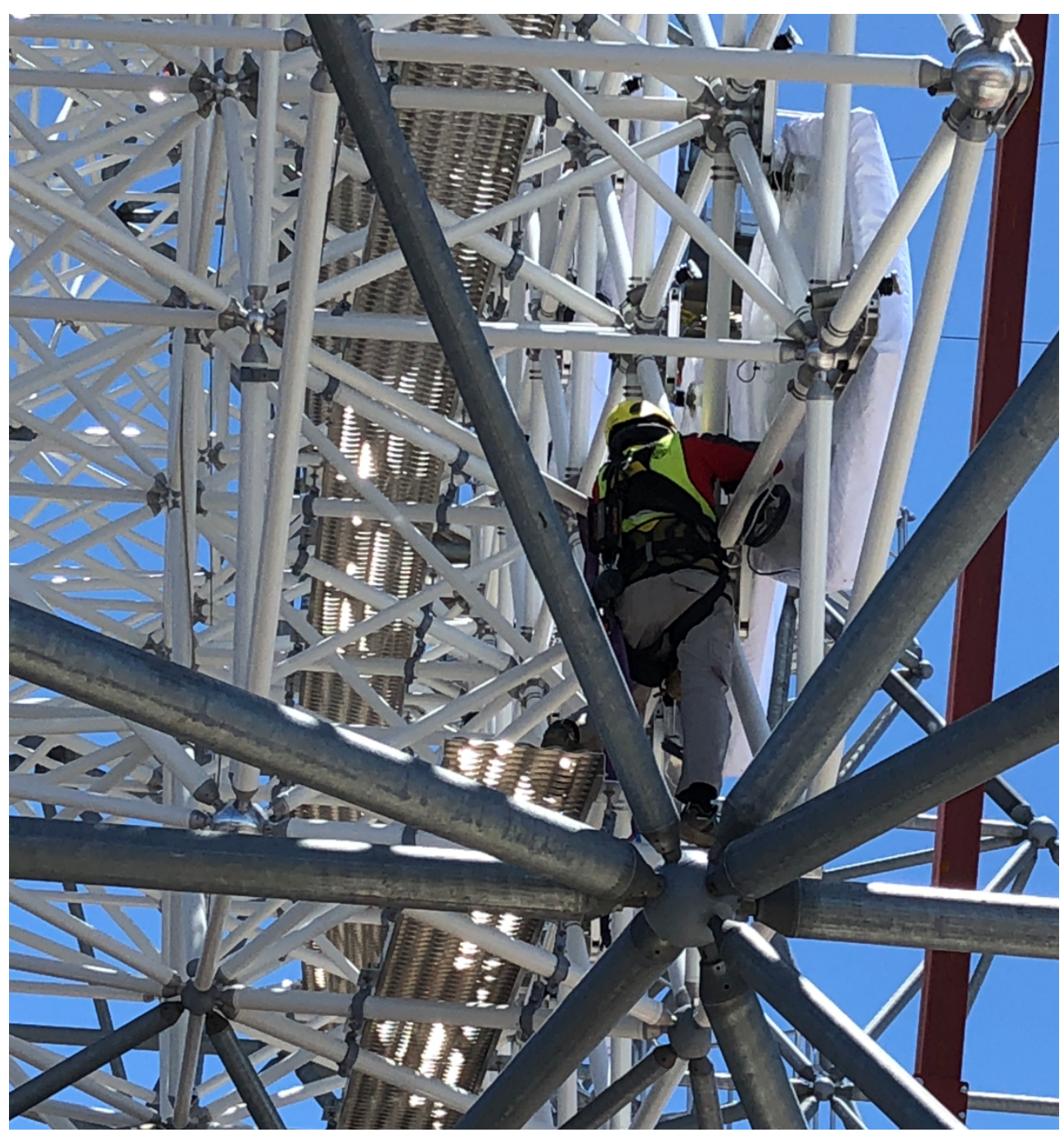

Figure 4: View into the dish while the mirrors are being integrated

The main structure is composed of several types and sizes of truss members. The predominant member type is a quasi-pinned strut system that allows the realization of arbitrary space-frame geometries while facilitating rapid assembly. These members are interconnected through standardized spherical nodes.

Both the struts and the nodes are available in various geometries and materials to meet the local stiffness and strength requirements while achieving an overall efficient structural configuration. Depending on the specific load and stiffness demands, the struts are manufactured from galvanized structural steel, aluminum alloys, or carbon-fiber-reinforced polymer (CFRP), with particular emphasis on minimizing structural mass.

The connections are considered quasi-pinned because the strut end fittings are relatively slender compared to the main member cross-sections. As a result, they exhibit a significantly lower rotational stiffness, allowing limited angular deformation and thereby approximating pinned-joint behaviors within the design load range

Table 3: Characteristics of the structs used for the space frame design of the LST

| **Struts in azimuth structure** | |
|---|---|
| ● Total number of steel struts | 390 |
| ● Total mass | 21500 kg |
| **Struts in the dish structure** | |
| ● Total number of CFRP struts | 2510 |
| ● Total number of aluminum struts | 162 |
| ● Total number of steel struts | 162 |
| ● Total mass | 18100 kg |

The second type of structural member consists of uniform-section struts terminated by flat flange connections, which provide significant bending stiffness in addition to axial load-carrying capacity. These members are primarily used at locations where rigid connections are required, such as the attachment of subsystems through dedicated nodes, including the azimuth bogies, maintenance platforms, and other auxiliary structural elements.

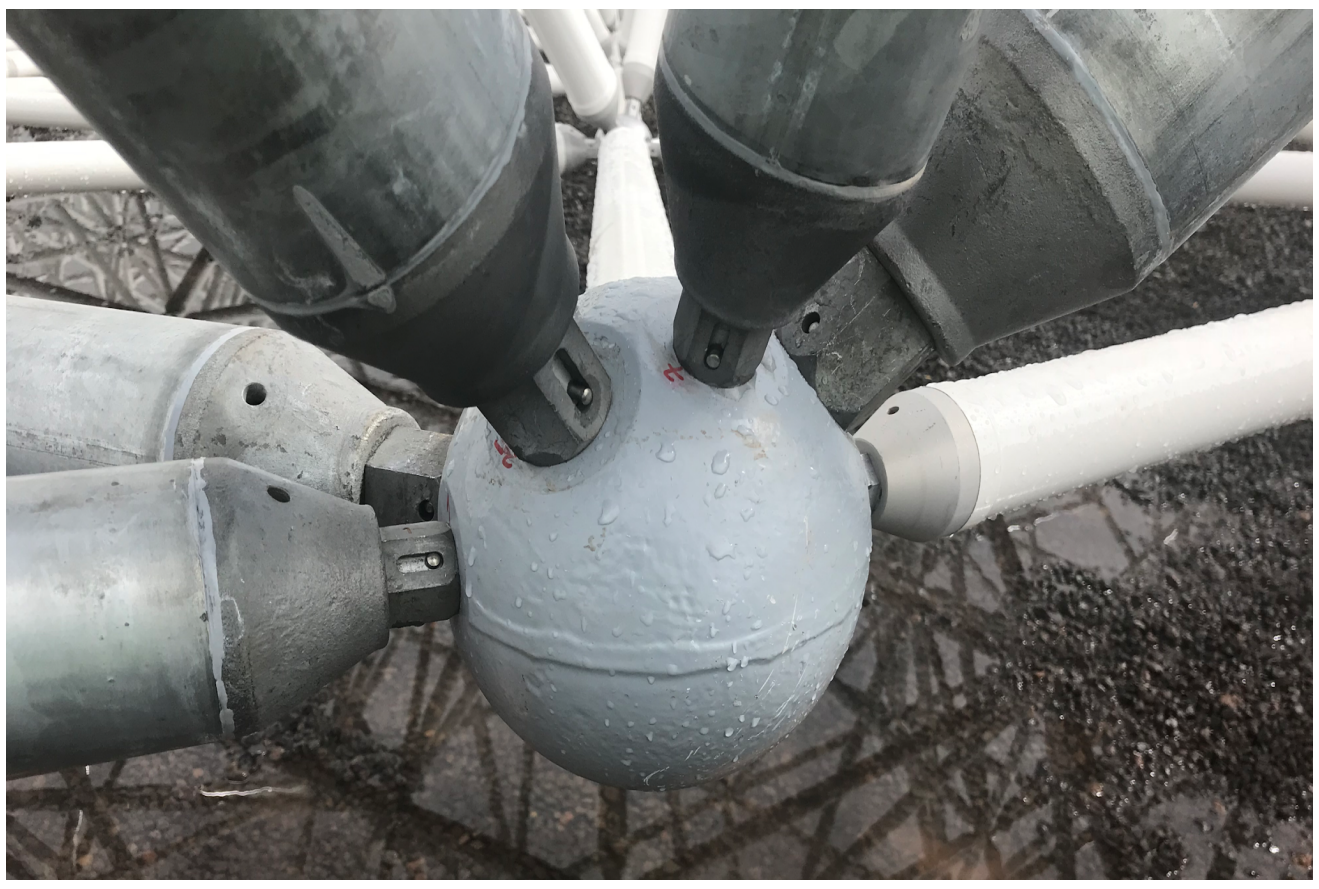

Figure 5: Standard MERO-TSK strut system (quasi-pinned)

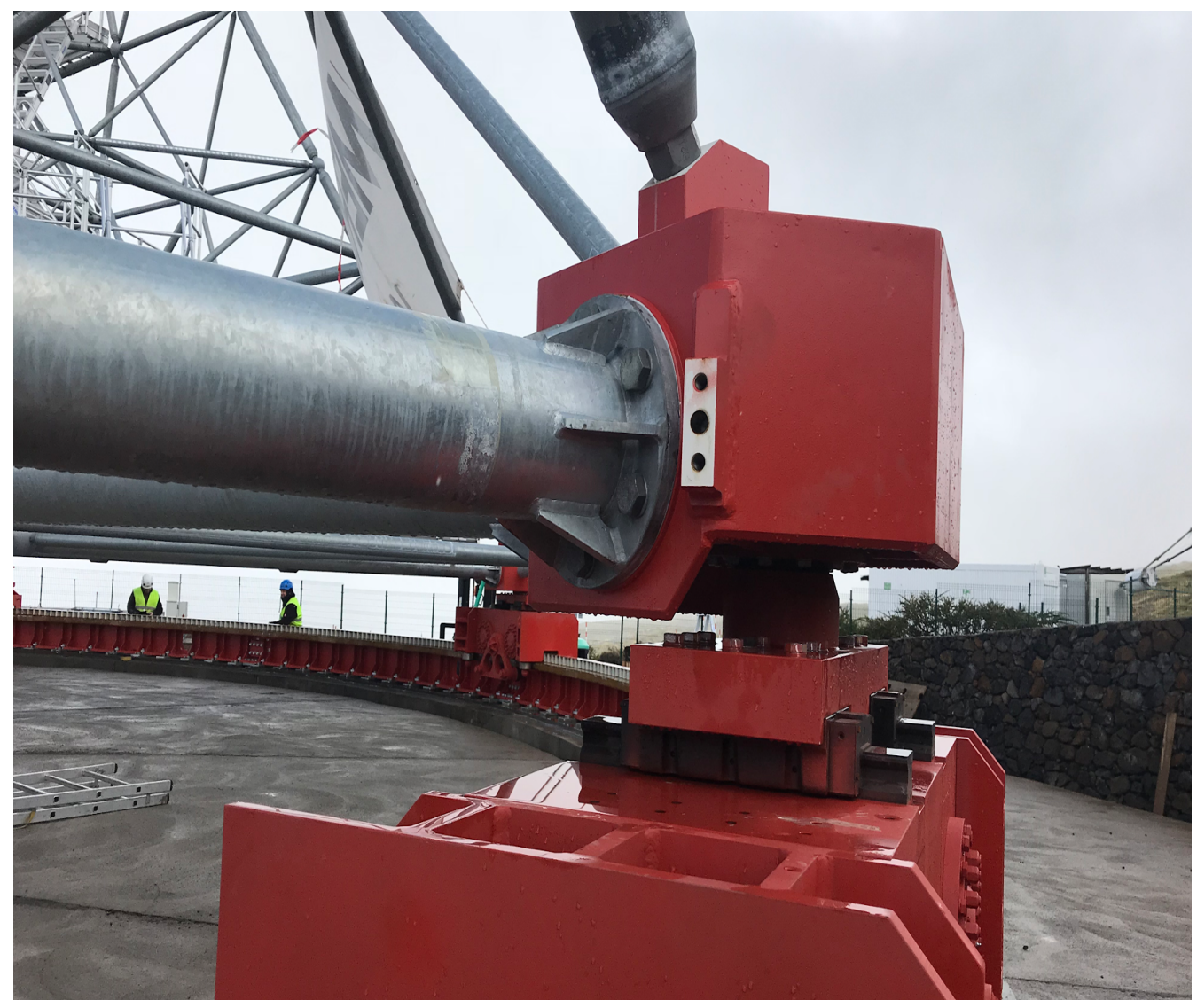

Figure 6: Firmly fixed struts with custom knot for AZ bogie attachment

The dish backup structure is driven in elevation through the rear arch, which is equipped with a duplex chain. The elevation drive unit is suspended from the rear arch and engages directly with the chain. The floating drive assembly is connected to the AZ structure by a reaction strut that transfers the drive forces into the telescope mount. The rear arch is stabilized by a system of steel tension cables and supports a set of counterweights used to balance the elevation assembly. This arrangement reduces the required drive torque and enables smooth and efficient telescope motion.

The dish is mounted on top of the AZ structure towers by means of large spherical bearings left and right. As shown in the following figure. The local struts are radially oriented towards the rotation axis and fixed to the spherical bearing head. The shaft is firmly fixed to a massive bearing block.

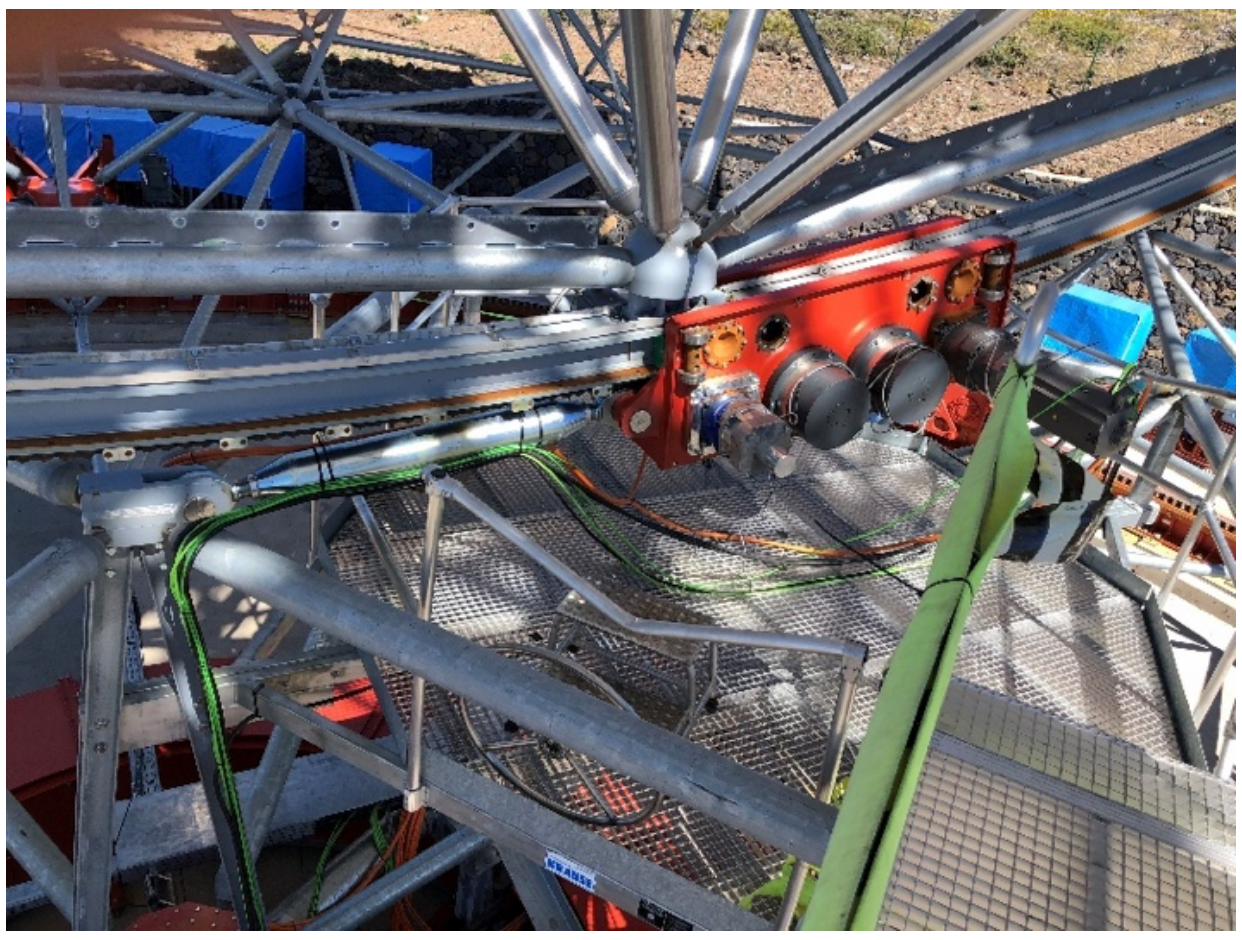

Figure 7: EL structure back arch with floating elevation drive linked to the AZ structure

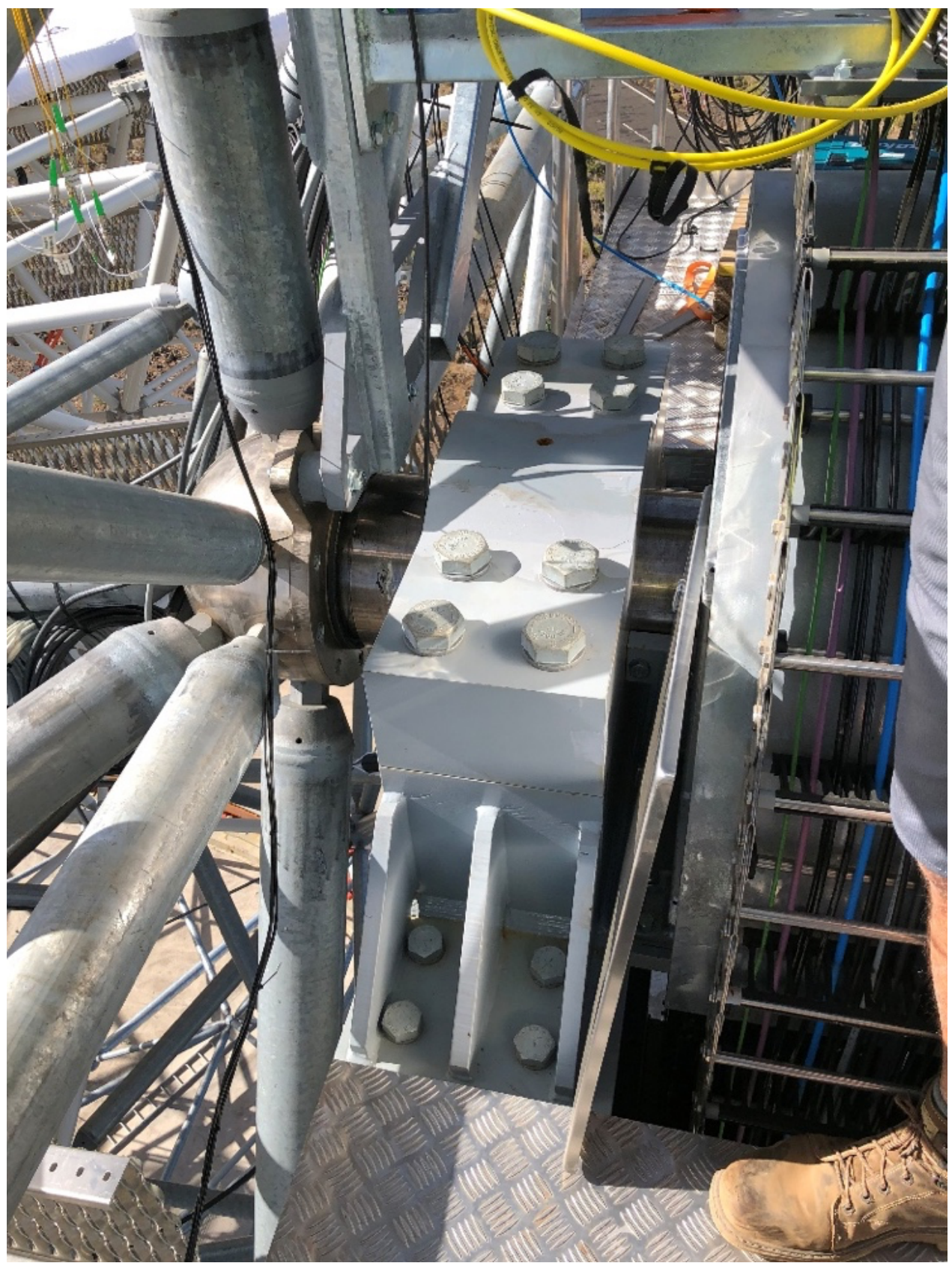

Figure 8: Elevation bearing with dish fixation on the left side and cable chain on the right side

Finally, the main structure is complemented by the Camera Support Structure (CSS). Figure 7 shows the complete telescope structure and highlights the different structural materials employed. The AZ structure is constructed entirely from steel, whereas the elevation structure utilizes a combination of materials to minimize total mass while providing a sufficiently stiff support for the mirror segments.

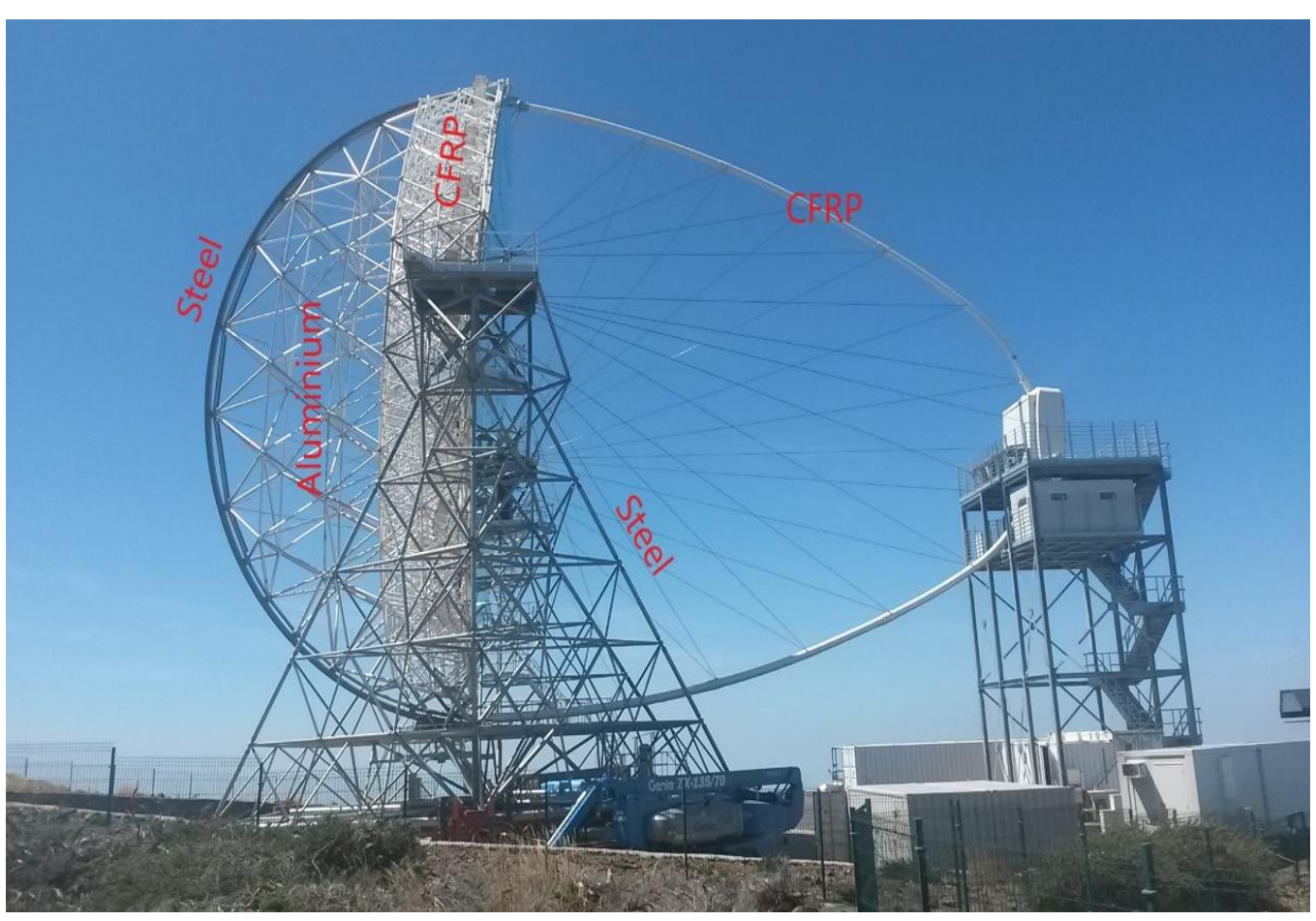

Figure 9: Telescope structure assembly in parking position with the CSS, tether and camera frame (furnished by LAPP)

The CSS consists of a thick-walled CFRP arch stabilized by a system of 26 CFRP tension members. These ropes are highly preloaded to maintain the arch in its nominal position and ensure that it remains permanently in compression under all operational conditions.

The CSS is manufactured entirely from CFRP to minimize its mass and thereby reduce the imbalance of the elevation assembly and the associated counterweight requirements on the rear arch. The CSS itself has a mass of approximately 2.5 tons; however, the dominant contribution to the elevation imbalance originates from the camera, which has a nominal mass of 2.5 tons and is located 28 m from the elevation axis.

To ensure a symmetric and balanced assembly of the CSS arch and tether system, a dedicated installation procedure was developed. The arch was initially supported in its upright position by a crane while the CFRP ropes were installed and sequentially tensioned. The preload in each rope was continuously monitored using load cells to achieve the specified force distribution. The resulting rope tensions are summarized in Table 4.

Table 4: Rope preloads versus gravity and elevation angle

| | EL ID | PRELOAD ONLY | EL = 0° PRELOAD + GRAV NO CAMERA | EL = 0° PRELOAD + GRAV | EL = 30° PRELOAD + GRAV | EL = 60° PRELOAD + GRAV | EL = 90° PRELOAD + GRAV | EL = 95° PARKED PRELOAD + GRAV |
|---|---|---|---|---|---|---|---|---|
| CSS ROPES | 900101 | 17500 | 13640 | 14054 | 12442 | 12120 | 13251 | 13571 |
| | 900102 | 15000 | 11661 | 12296 | 10607 | 10039 | 10804 | 11058 |
| | 900103 | 13500 | 10299 | 11029 | 9495 | 8983 | 9684 | 9916 |
| | 900104 | 18500 | 13756 | 14733 | 12987 | 12646 | 13881 | 14232 |
| | 900105 | 33500 | 25479 | 26301 | 23553 | 23345 | 25863 | 26525 |
| | 900106 | 38000 | 28966 | 26137 | 24452 | 26265 | 31232 | 32299 |
| | 900107 | 31000 | 20283 | 17195 | 18922 | 23750 | 30526 | 31738 |
| | 900108 | 38000 | 28871 | 26013 | 30643 | 37112 | 43829 | 44859 |
| | 900109 | 33500 | 25424 | 26234 | 30713 | 35816 | 40312 | 40929 |
| | 900110 | 18500 | 13735 | 14722 | 17365 | 20234 | 22639 | 22952 |
| | 900111 | 13500 | 10309 | 11044 | 13153 | 15302 | 16967 | 17162 |
| | 900112 | 15000 | 11676 | 12311 | 14624 | 16977 | 18788 | 19007 |
| | 900113 | 17500 | 13654 | 14067 | 16508 | 19173 | 21402 | 21788 |
| | 900201 | 17500 | 13638 | 14052 | 12419 | 12081 | 13208 | 13528 |
| | 900202 | 15000 | 11661 | 12296 | 10609 | 10043 | 10807 | 11062 |
| | 900203 | 13500 | 10299 | 11029 | 9497 | 8987 | 9689 | 9922 |
| | 900204 | 18500 | 13755 | 14732 | 12992 | 12656 | 13893 | 14244 |
| | 900205 | 33500 | 25481 | 26302 | 23548 | 23336 | 25851 | 26513 |
| | 900206 | 38000 | 28972 | 26143 | 24454 | 26263 | 31226 | 32292 |
| | 900207 | 31000 | 20272 | 17184 | 18913 | 23744 | 30525 | 31738 |
| | 900208 | 38000 | 28875 | 26018 | 30651 | 37121 | 43836 | 44866 |
| | 900209 | 33500 | 25426 | 26236 | 30720 | 35827 | 40324 | 40941 |
| | 900210 | 18500 | 13734 | 14721 | 17358 | 20223 | 22626 | 22939 |
| | 900211 | 13500 | 10308 | 11043 | 13150 | 15298 | 16963 | 17157 |
| | 900212 | 15000 | 11676 | 12312 | 14622 | 16975 | 18785 | 19004 |
| | 900213 | 17500 | 13652 | 14066 | 16525 | 19205 | 21438 | 21824 |
| DISH ROPES | 950101 | 150000 | 161915 | 162890 | 159263 | 153440 | 146582 | 145459 |
| | 950102 | 150000 | 166596 | 167774 | 164359 | 157766 | 149008 | 147487 |
| | 950103 | 150000 | 166197 | 167374 | 166231 | 160880 | 152706 | 151198 |
| | 950104 | 150000 | 161771 | 162745 | 163988 | 160524 | 154950 | 153860 |
| | 950201 | 150000 | 162007 | 162994 | 159267 | 153385 | 146488 | 145360 |
| | 950202 | 150000 | 166614 | 167793 | 164343 | 157730 | 148952 | 147429 |
| | 950203 | 150000 | 166198 | 167378 | 166315 | 160964 | 152771 | 151258 |
| | 950204 | 150000 | 161821 | 162808 | 164096 | 160682 | 155074 | 153977 |

All drive systems are equipped with brakes capable of holding the telescope under moderate wind conditions, while the bogies are provided with anti-uplift devices to prevent separation from the azimuth rail. In anticipation of extreme storm or hurricane-force winds, the telescope can be placed in a locked configuration to prevent unintended rotation caused by asymmetric wind loading that could exceed the holding capacity of the drive systems.

The elevation axis is secured by a redundant locking system consisting of locking pins that engage directly within the elevation drive mechanism, as shown in Figure 6. Rotation about the azimuth axis is prevented by an azimuth locking system developed by IFAE. This system acts on the two driven bogies and clamps them to the foundation. Functionally, the mechanism operates like a pair of pliers, gripping the azimuth node located above each driven bogie and thereby preventing rotation. The azimuth locking system is shown in Figure 15.

To limit the motion of the 2.5-ton camera located 28 m from the elevation axis, a dedicated camera locking system is provided. This system is mounted on the access tower directly beneath the camera and engages the arch of the Camera Support Structure (CSS), thereby reducing camera motion and the associated loads during severe environmental conditions. The camera locking system is shown in Figure 10.

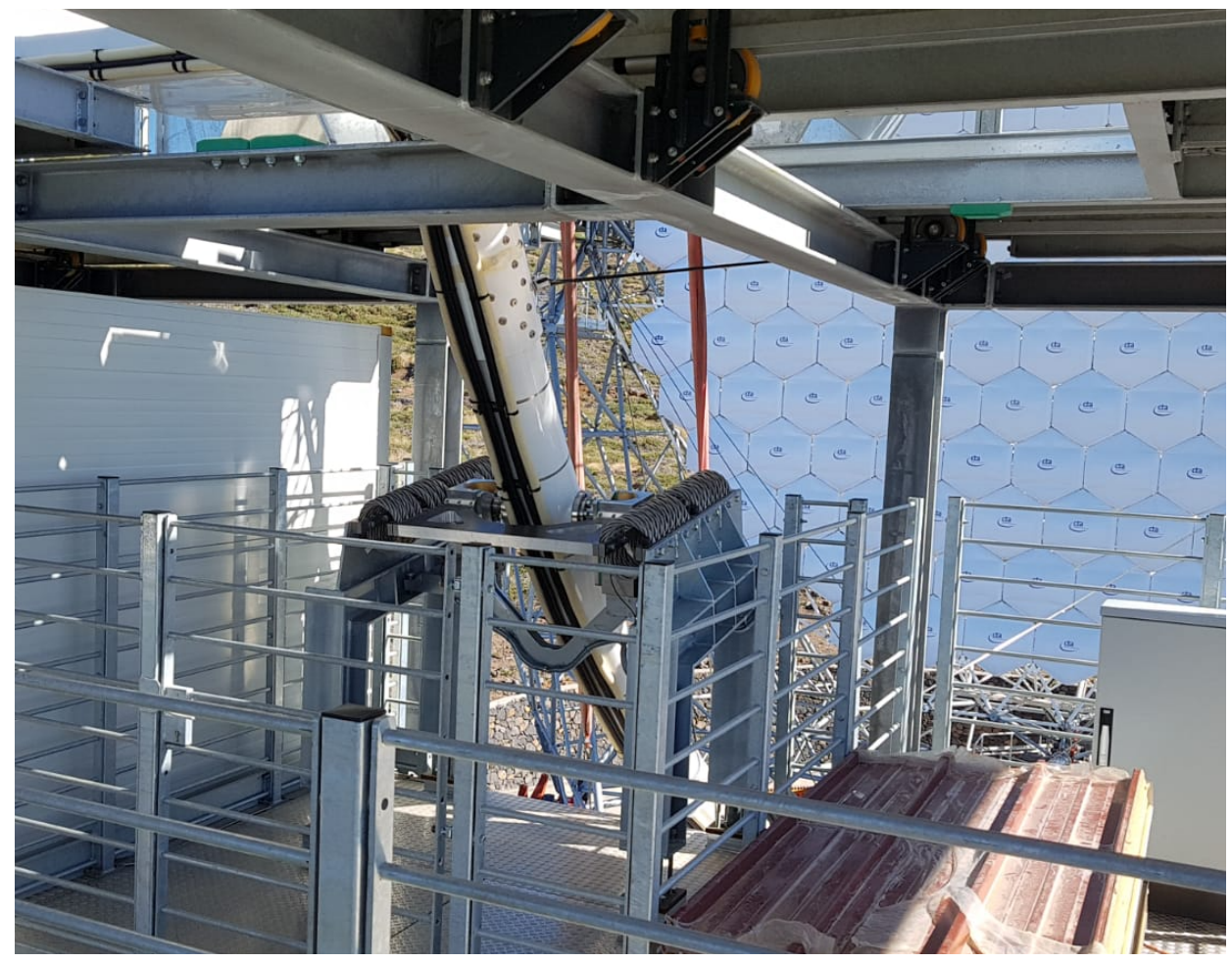

Figure 10: Camera locking system mounted on the access tower (the primary mirror in the background)

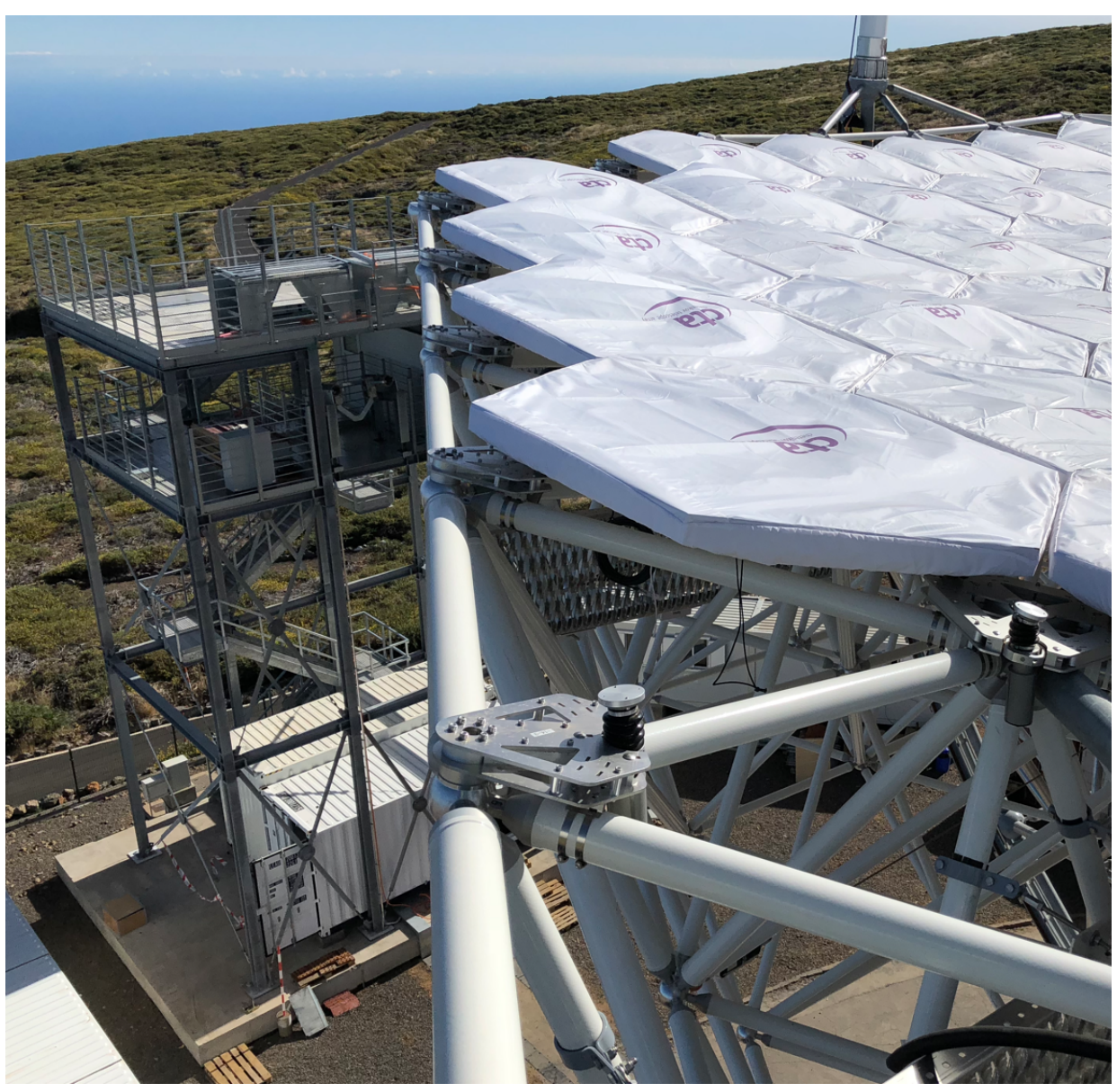

Figure 11: Mirror segment mounting to active mirror control units

The mirrors are fixed to the dish by means of the active mirror control system (AMC) mounted to the structural knots of the dish CFRP struts. Each AMC unit consists of three actuators facilitating piston and tilt correction of the individual mirrors. The following figure shows the typical mirror segment interface.

## 4. STRUCTURAL CHARACTERISTICS

The structural characteristics of the telescope were determined using a detailed NASTRAN finite-element model of the main structure, including the Camera Support Structure (CSS). The various subsystems were represented by equivalent mass and stiffness models where appropriate. The finite-element model primarily captures the global structural behavior and includes the principal load-carrying elements, such as the truss members, arches, and tension ropes.

The analysis focused on the global stiffness, strength, and dynamic characteristics of the telescope structure. Local interface details and subsystem components were not modelled explicitly; instead, they were assessed separately using the interface loads, displacements, and accelerations derived from the global structural model. This approach ensured a consistent transfer of load and dynamic response information between the overall telescope structure and the individual subsystem designs.

The relevant conditions are the following. A brief summary of the FEM results are provided in the following pages.

- Gravity under different elevation angles and preloads
- Max. wind gusts at 50 km/h during operations
- Max. wind gust at 83.5 km/h during emergency parking
- Max. wind gust at 200 km/h in parked/safe position
- Ambient temperature, solar radiation and reflected sun light
- Accumulated ice from dedicated ice storms

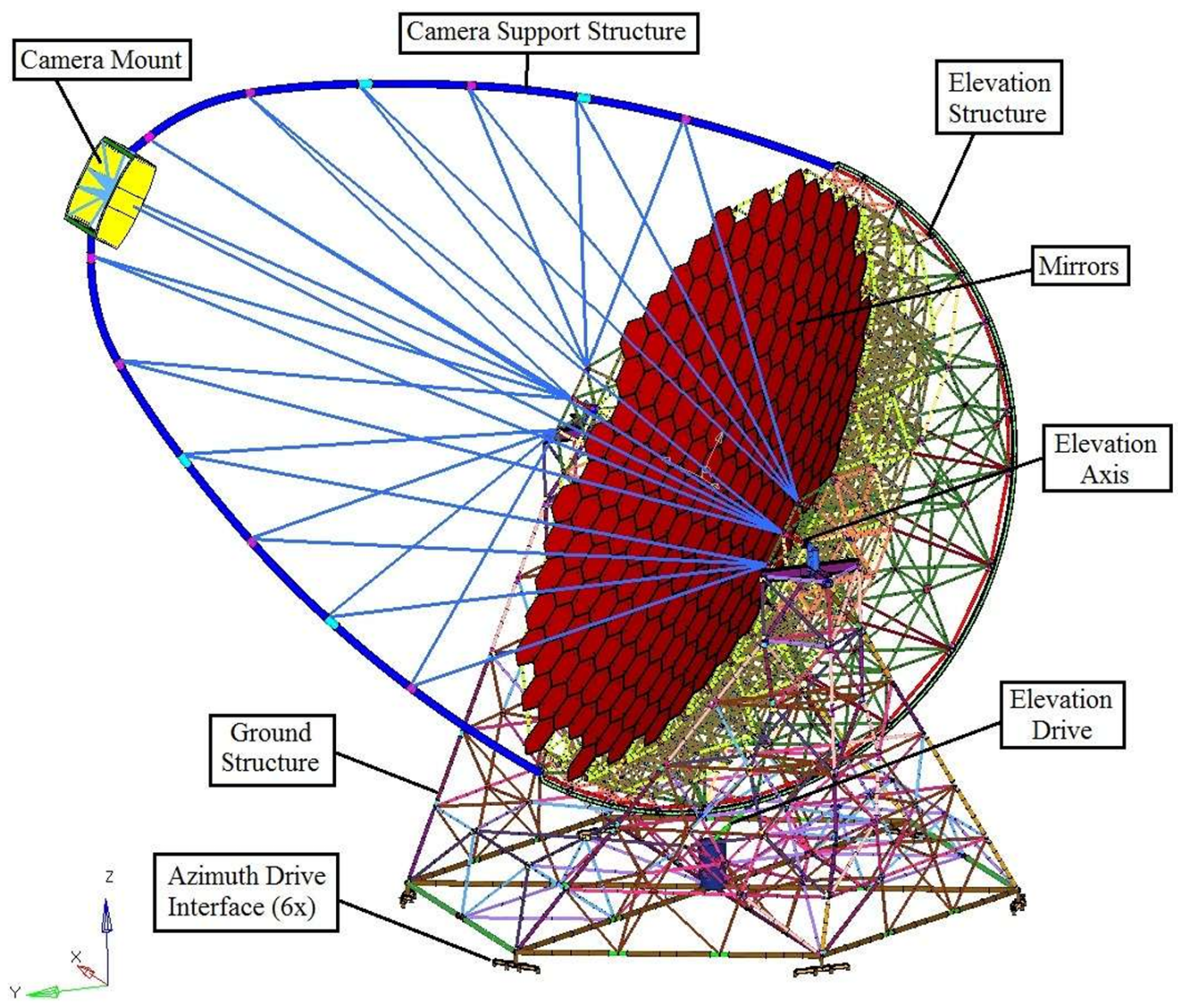

Figure 12: Main structure FEM model

### 4.1 Structure deformation

The following figures qualitatively illustrate the structural stiffness and deformation of the telescope under gravity and wind loading for both operational and parking (survival) configurations. The displayed deformation amplitudes represent the global structural displacements and cannot be directly interpreted as optical misalignment errors, which require additional post-processing and optical performance analysis.

Table 5 summarizes the relevant deformation parameters for the various load-case combinations considered in the structural assessment:

Table 5: Peak displacements of the main structure under static loads

| Load Case | Elevation (deg) | Camera (mm) | Max (mm) |
|---|---|---|---|
| Preload only | 0 | 5.3 | 41 |
| Gravity only | 0 | 22.4 | 23 |
| PL+GRAV+TEMP+WIND83.5 (limit load) | 0 | 30.4 | 44 |
| | 30 | 38.5 | 57 |
| | 60 | 45.9 | 71 |
| | 90 | 49.1 | 79 |
| PL+GRAV+TEMP+WIND200 (limit load) | 95P | 119.9 | 157 |
| PL+GRAV+TEMP+ICE+WIND120 (limit load) | 95P | 66.4 | 92 |

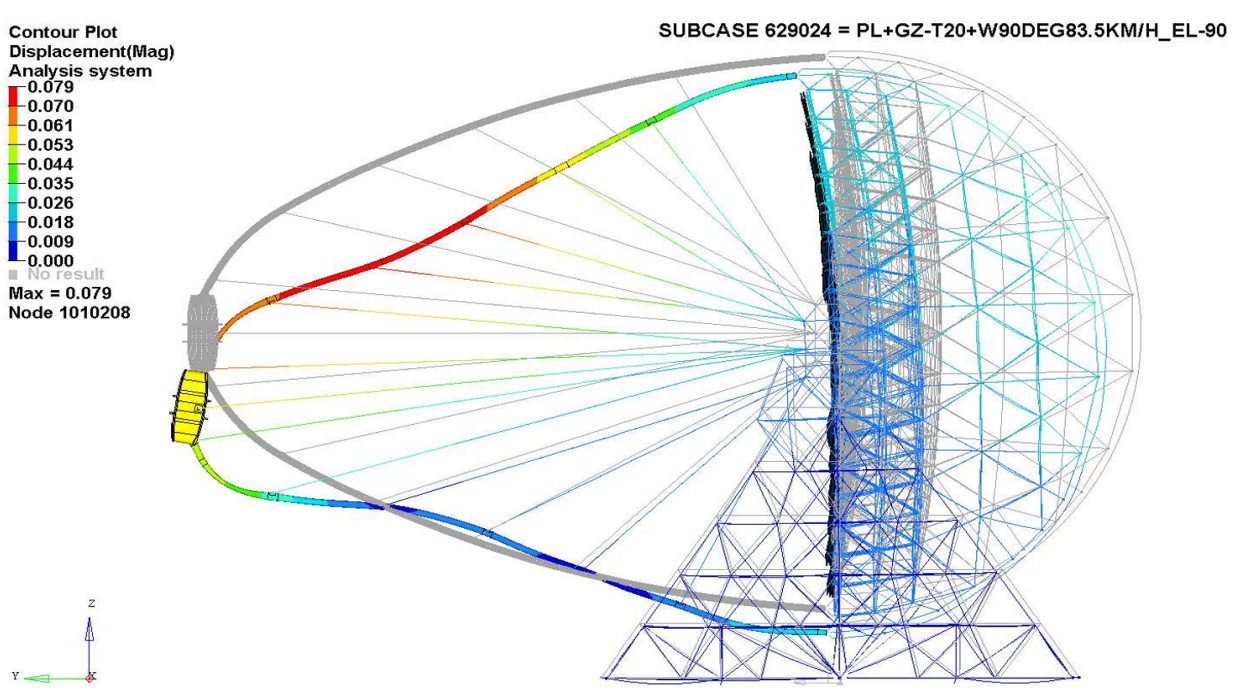

Figure 13: Qualitative deformation under gravity and back wind at 83.5 km/h for 90deg elevation (max. 79 mm)

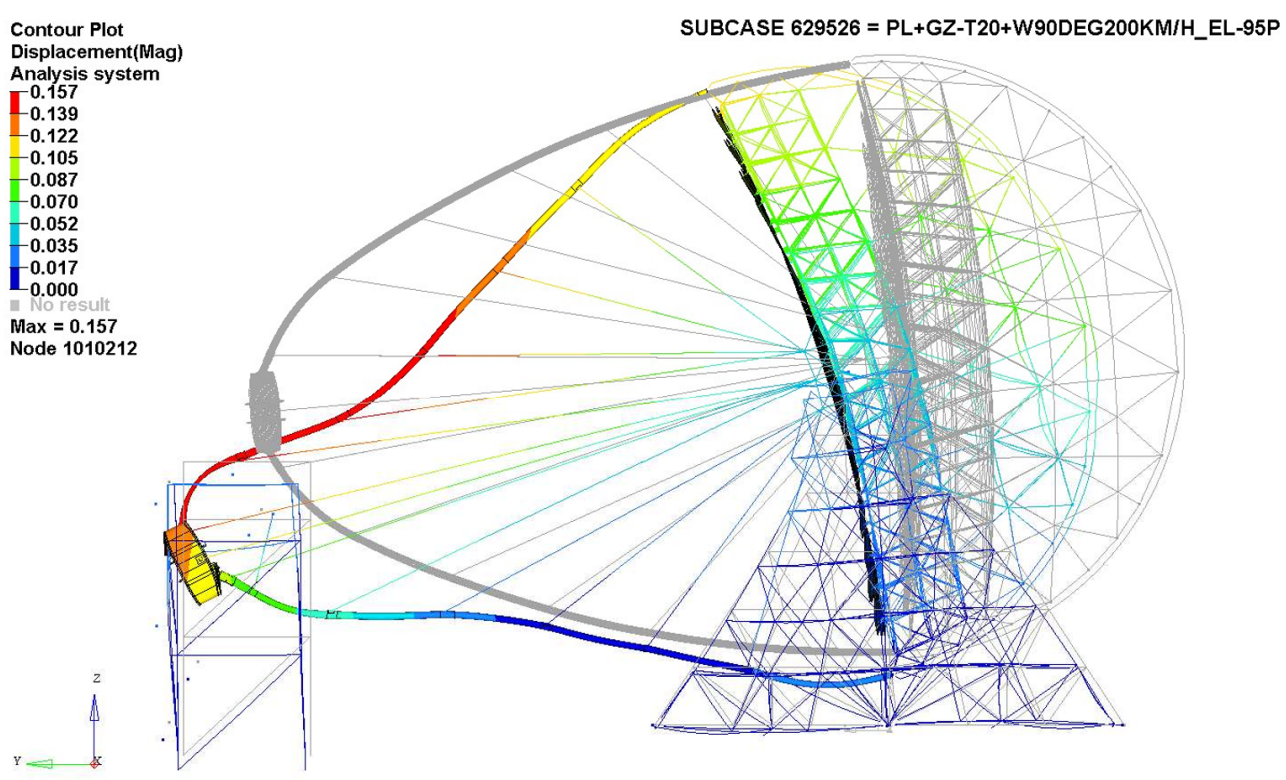

Figure 14: Qualitative deformation under gravity and back wind at 200 km/h in parking position (max. 157 mm)

### 4.2. Alignment stability

The alignment stability relevant to the optical pointing performance is determined by the relative displacements between the mirror segments and the focal-plane sensors located in the camera. Table X summarizes these relative displacements in terms of decentering along the two principal directions of the focal plane and defocus perpendicular to the focal plane. The analysis assumes that the telescope optics are aligned at the zenith position.

Whenever the resulting misalignment exceeds the specified optical tolerances, the Active Mirror Control (AMC) system can be used to compensate for the deviations. Consequently, it is only necessary to verify that the AMC actuators provide sufficient adjustment range to accommodate the required corrections under all operational conditions.

Table 6: Mis-alignment due to elevation change and wind

| Elevation | Load case | De-Centre X (mm) | De-Centre Y (mm) | De-Focus (mm) |
|---|---|---|---|---|
| 0° (zenit) | Gravity | 0 | 0 | 0 |
| | Steady wind | 7.23 | 0.06 | 0.34 |
| | Dynamic wind | 4.79 | 0.22 | 0.06 |
| 60° | Gravity | 10.6 | 22.4 | -12.6 |
| | Steady wind | 6.17 | 0.31 | 0.06 |
| | Dynamic wind | 3.99 | 0.07 | 0.06 |

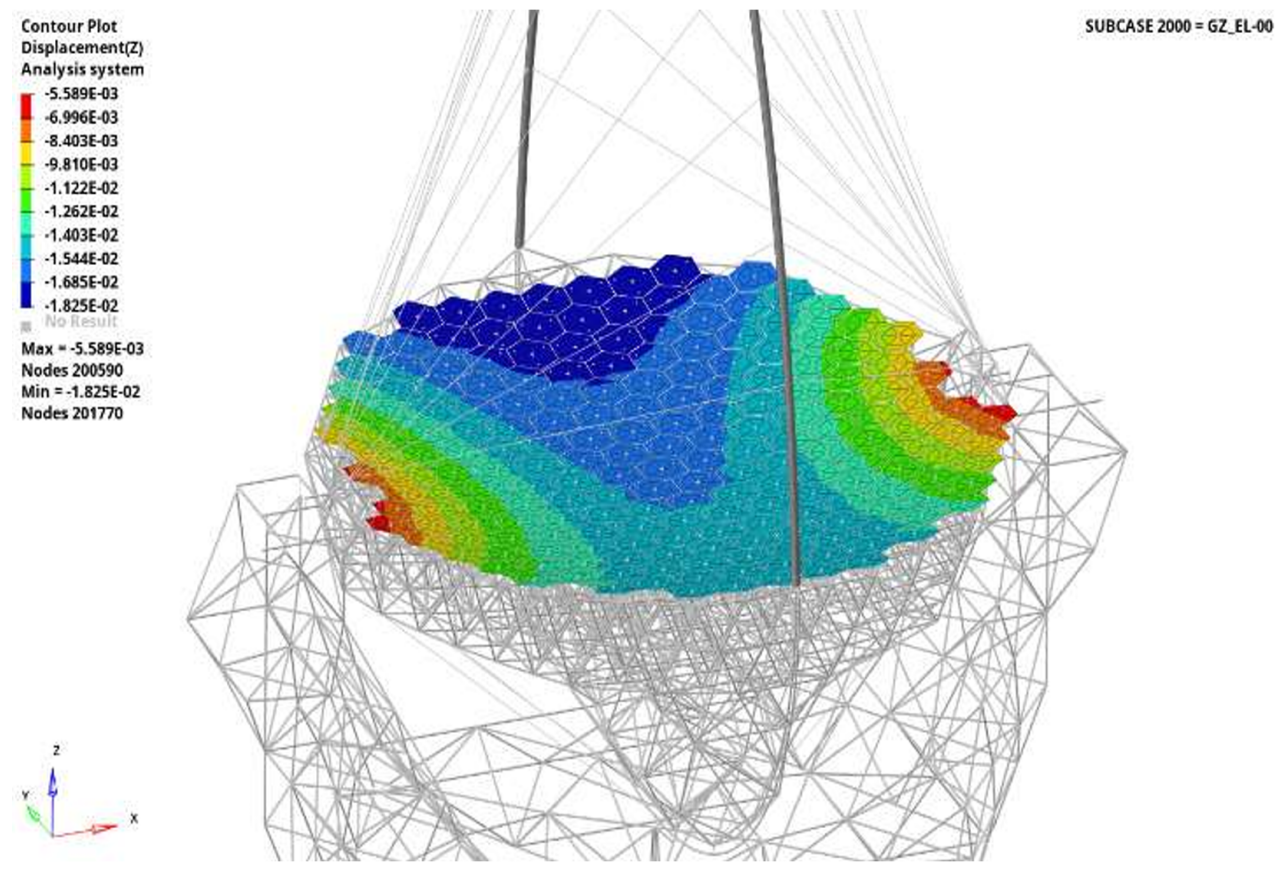

Figure 15: Sag of the dish under gravity (11 mm PTV)

### 4.3 Dynamic characteristics

The natural vibration modes are of particular importance for alignment stability under dynamic wind loading, for smooth telescope slewing, and for the structural response to seismic excitation. The following figures illustrate the characteristic deformation patterns associated with the lowest eigenmodes of the telescope structure, which are governed by its stiffness and mass distribution. These mode shapes provide valuable insight into the global dynamic behavior of the telescope and identify the dominant mechanisms contributing to structural deformation under dynamic loading.

Table 7: Fundamental eigenfrequencies

| Elevation | 0 deg | 30 deg | 60 deg | 90 deg | 95 deg |
|---|---|---|---|---|---|
| Lateral mode along elevation axis | 1.91 | 1.90 | 1.90 | 1.90 | 2.05 |
| Lateral mode perpendicular to elevation axis | 2.23 | 2.19 | 2.11 | 2.05 | 2.39 |
| Vertical mode along Zenith axis | 3.98 | 3.89 | 3.80 | 3.67 | 4.08 |

### 4.4 Loads

The telescope is designed to resists the following operational and environmental conditions:

- Azimuth and elevation drive max angular acceleration 100 mrad/s2
- Survival air temperature range -20°C to +40°C
- Max. short term gust in safe state condition 200 km/h
- Max. layer of ice in safe state condition 20 mm
- Seismic accelerations (PGA) 0.05 g

#### 4.4.1 Wind loads

Wind loading is the dominant design load for the telescope, generating forces exceeding 100 tons on the primary reflector. Owing to the lightweight design of the structure, the telescope's self-weight alone is insufficient to maintain contact between the bogies and the azimuth rail at wind speeds of approximately 100 km/h and above. To prevent uplift, all bogies are equipped with anti-uplift devices that engage beneath the rail running surface.

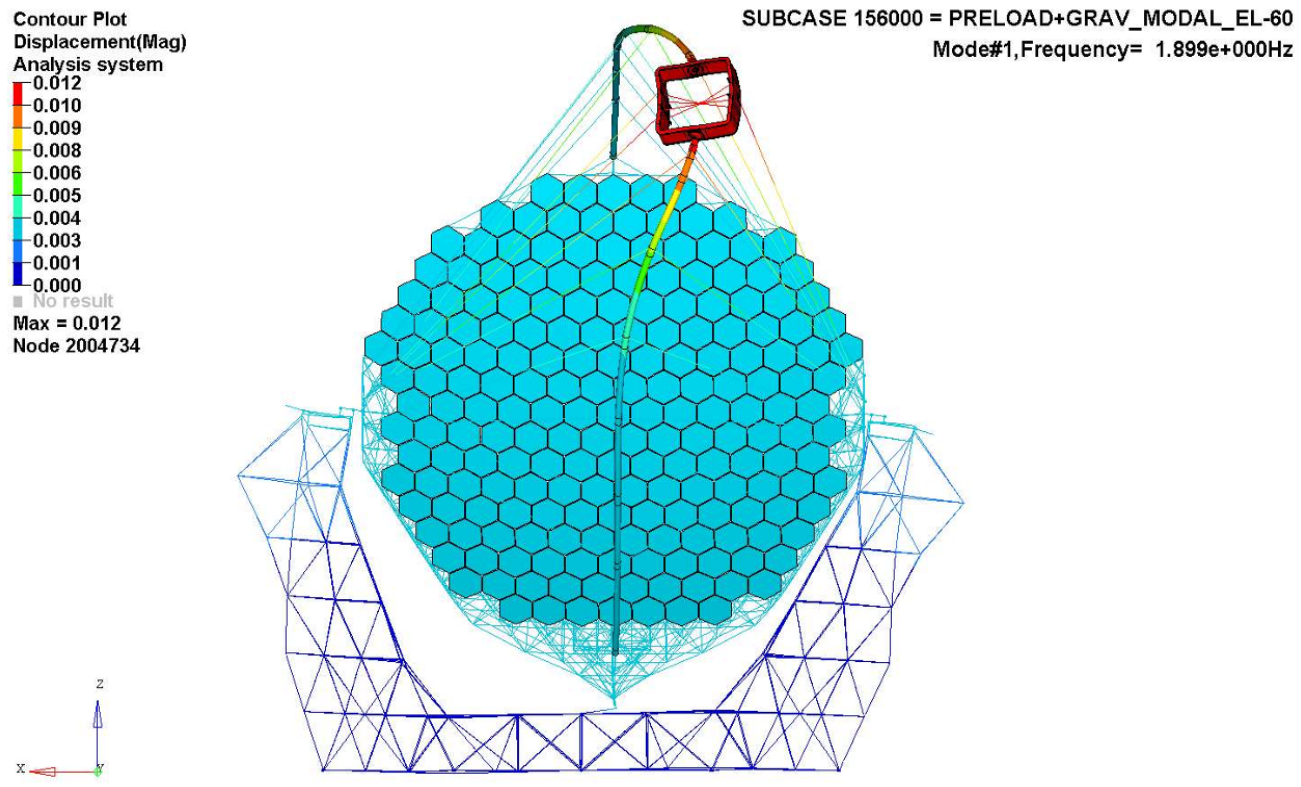

Figure 16 : Mode 1 at 1.889 Hz for 60deg elevation,

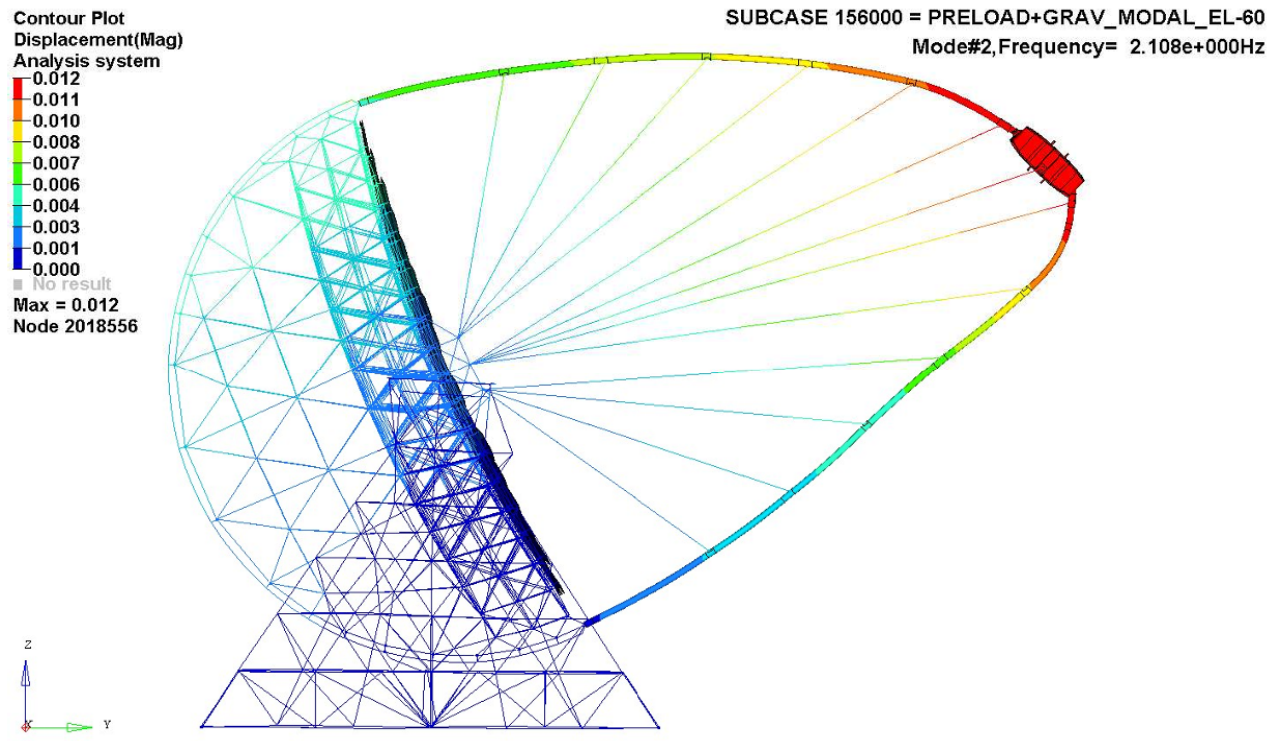

Figure 17: Mode 2 at 2.108 Hz for 60deg elevation

To ensure safe operation, the telescope is automatically moved to its parking position when the average wind speed exceeds 50 km/h. Furthermore, the azimuth drive system is not designed to withstand the large asymmetric wind moments acting on the reflector under survival conditions. Therefore, two dedicated locking devices are provided at the driven bogies, where they clamp structural nodes of the telescope mount and transfer the resulting wind loads directly into the azimuth structure and foundation.

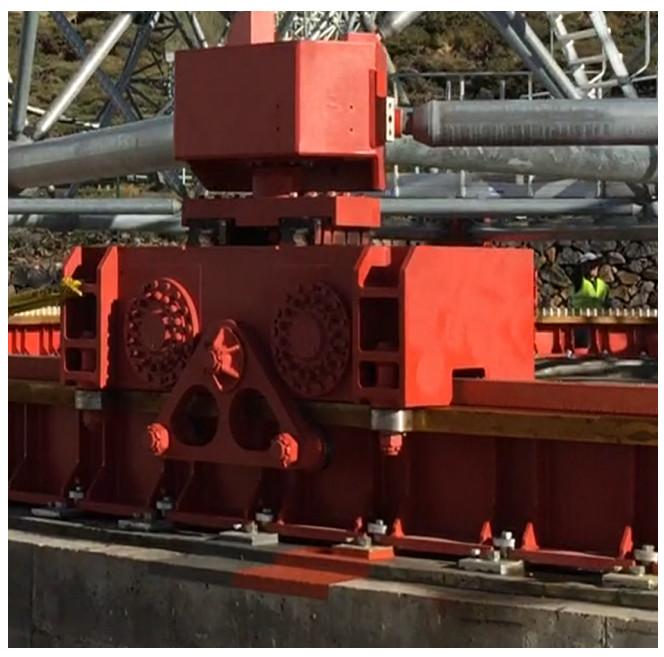

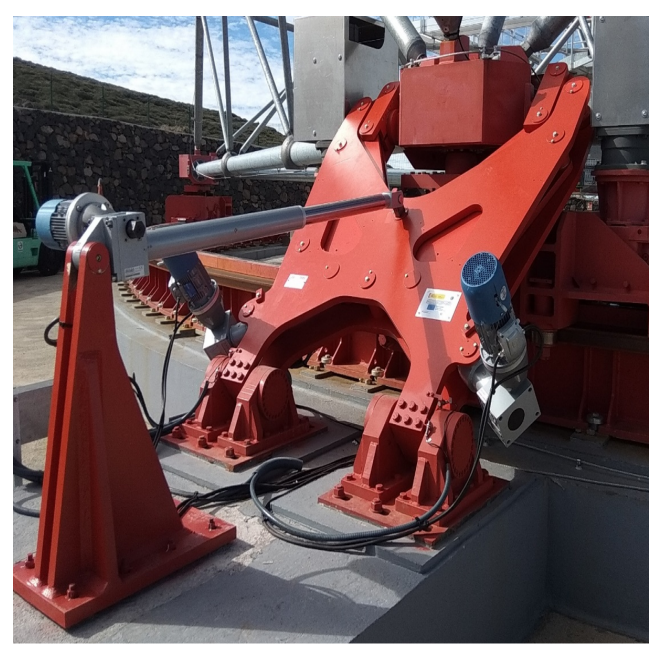

Figure 18: Retention of the telescope on the azimuth system designed by IFAE (left: anti-uplift, right: rotation)

**4.4.2 Oscillation of tubes due to vortex shedding**

Another important aspect of the structural design is the susceptibility of some of the slender aluminium struts located on the rear side of the dish to vortex-induced vibrations. Although the structural supplier, MERO-TSK, was well aware of this phenomenon and had established critical length-to-diameter limits based on previous analyses and experimental investigations, the failure of a single strut occurred approximately one year after commissioning. In response, flow-disturbing devices were installed on a number of struts identified as being close to the critical geometric limits for vortex excitation. Subsequently, a comprehensive assessment was carried out to identify all potentially critical members within the structure, comprising nearly 3,000 individual struts.

A detailed fatigue assessment revealed an additional life-limiting component associated with repeated elevation movements of the telescope. The critical location is the joint of a short structural member connected to the elevation axis and not affected by vortex-induced vibrations. The maximum tensile load occurs at horizon pointing, whereas the minimum load is reached at zenith pointing, resulting in a significant cyclic stress range during normal operation. The fatigue analysis demonstrated that an excessive number of elevation cycles could ultimately limit the service life of this connection.

To verify the fatigue resistance of the critical joint, dedicated fatigue tests were performed on the M27 connecting bolts, which were cycled between tensile loads of 100 kN and 200 kN. In addition, an adjacent member incorporating an M48 bolt and providing a fail-safe load path was tested under cyclic loads with a maximum tension of 500 kN. A comprehensive fail-safe assessment was subsequently conducted for both the local connection and the overall truss structure. The analysis demonstrated that the vast majority of structural members possess alternative load paths and can therefore be considered fail-safe. Likewise, the failure of individual CSS tension ropes was shown not to compromise the structural integrity of the telescope.
To monitor the accumulated fatigue damage throughout the operational lifetime of the telescope, a cycle-counting system has been implemented to record elevation movements and provide a continuous assessment of fatigue usage.

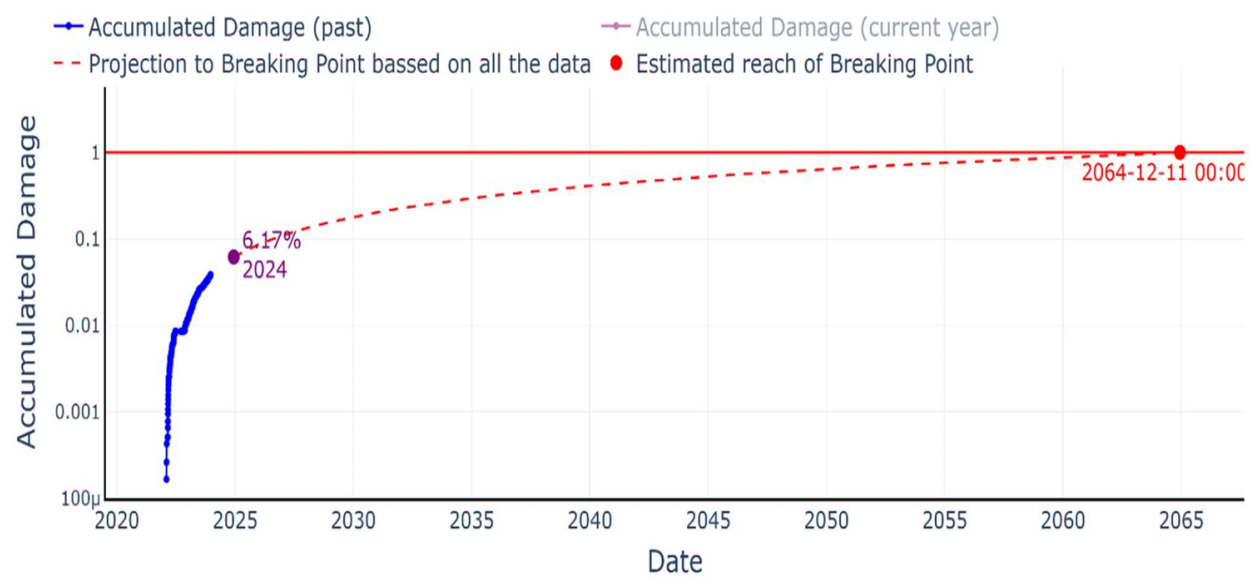

Figure 19: Simplified monitoring system for accumulated fatigue damage (Universityof Jaen)

### 4.5 Ice Loads During Winter

During the first winter on site, while the telescope was still only partially assembled, an ice storm was encountered that exceeded expectations, although similar events had previously been observed on the MAGIC telescopes. For the structural design, ice accretion was modelled as an additional gravitational load corresponding to approximately 40 tons of extra mass, representing severe icing conditions that had already been recorded on two occasions under exceptional environmental circumstances.

The telescope structure was designed to withstand this loading condition, which corresponds to an ice layer of approximately 20 mm thickness uniformly accumulated around all exposed structural members. The resulting ice load of approximately 40 tons was included in the structural verification of the telescope under survival conditions.

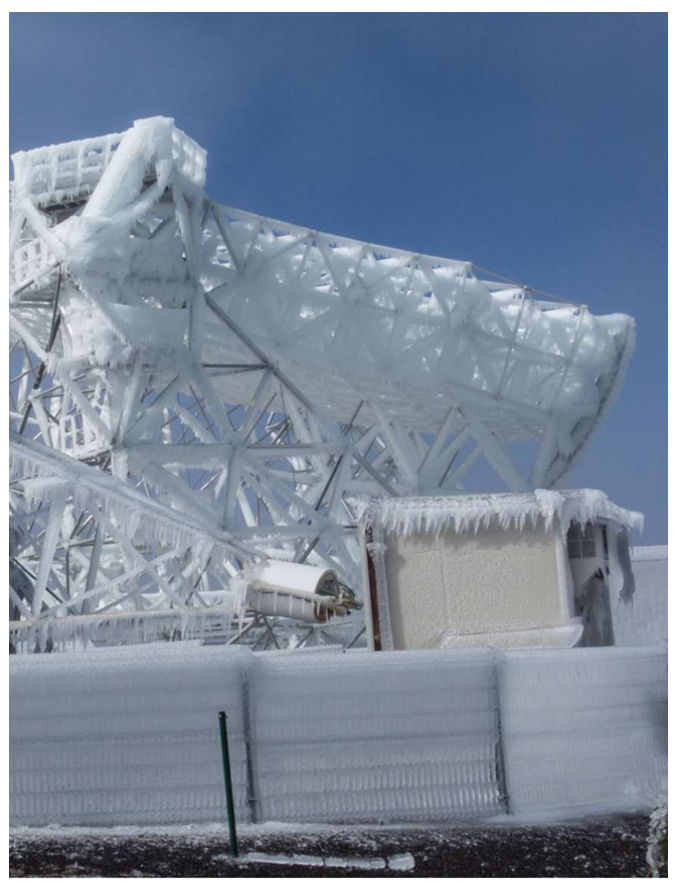

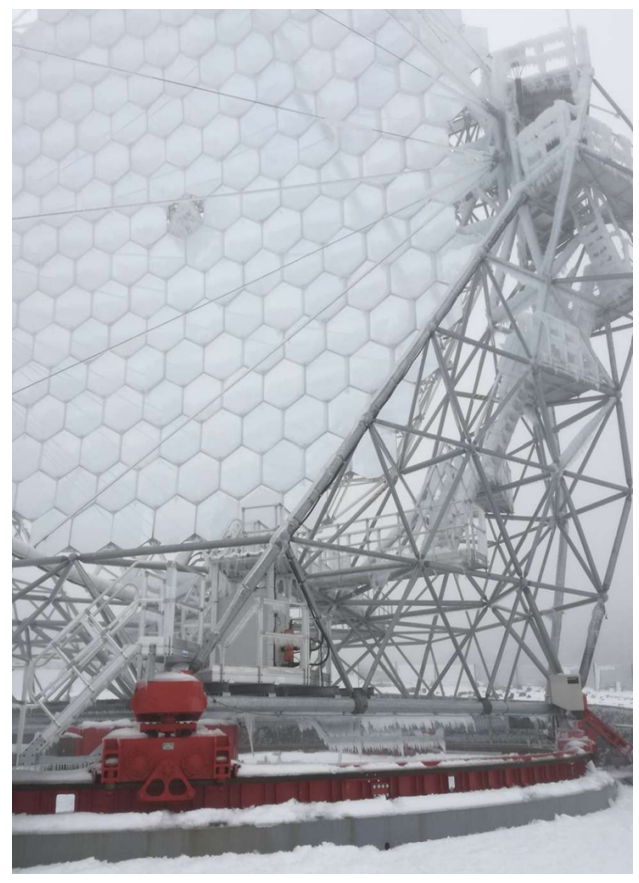

Figure 20: Ice loads 2018 during assembly (left)
and 2022 during commissioning (right)

Operational experience has shown that the additional weight of accumulated ice does not pose a critical challenge to the primary structure. A more significant concern is the detachment and subsequent fall of large ice blocks, which can damage secondary structures such as stairs, walkways, and maintenance platforms. Although such damage does not compromise the structural integrity or safety of the telescope, it requires periodic inspections and, when necessary, repair of the affected components following severe icing events.

### 4.6 Solar Radiation Reflected by the Reflector

Another important lesson learned during operation was the effect of sunlight reflected by the mirror segments. During daytime, the telescope is parked with the reflector pointing north and inclined approximately 5° below the horizon. It was already well understood from design studies that, around the summer solstice, sunlight could illuminate the reflector from the side, resulting in concentrated reflections focused onto specific locations in the surrounding environment. Potential impact zones and associated hazards had therefore been identified and assessed in advance.

Nevertheless, it was unexpectedly observed that reflected sunlight could also be focused onto the attachment region of the CSS tether closest to the dish. This effect had not been identified during the original analysis and demonstrated the complexity of predicting all possible reflection paths generated by the large segmented reflector. The localized concentration of solar radiation resulted in elevated temperatures at the affected components and required the implementation of additional protective measures.

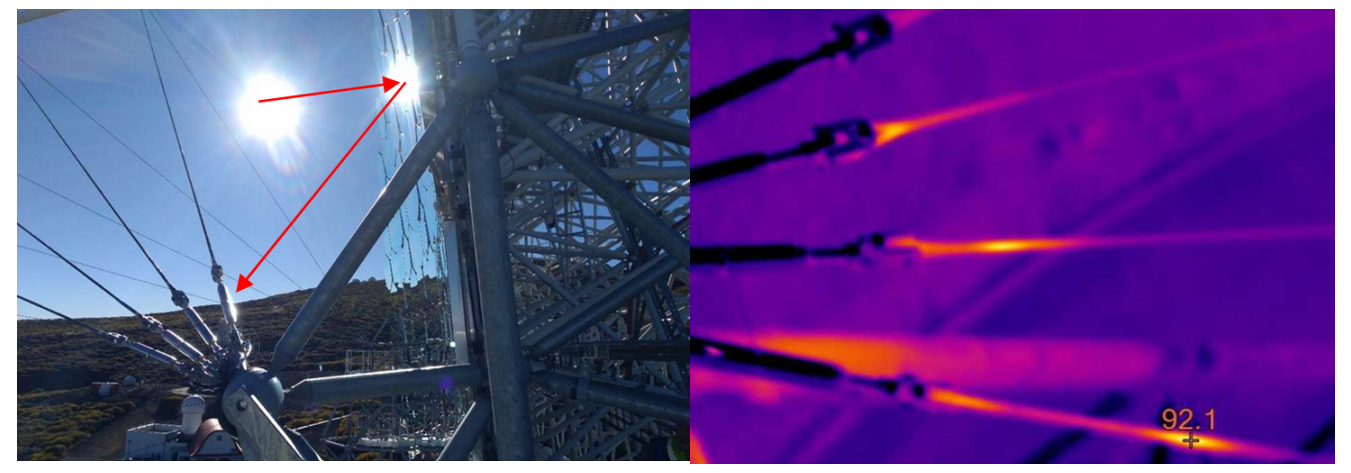

Figure 21: Excessive heating of the rope attachment by reflected sun light

This was leading to temperatures beyond 100 °C on the CFRP ropes and their mounting. Luckily the ropes and parts were selected to sustain the heat load by their specification. For mitigation of the related risk, thermal insulation was applied in the critical region to reduce the temperature of the ropes and corresponding turnbuckles.

### 4.7 Seismic loads in La Palma

During the early design phase, a seismic design scenario representative of the conditions at Paranal, Chile, was adopted. This conservative approach was motivated both by the limited seismic information available for La Palma at the time and by the possibility of deploying the LST design at a future CTA site in Chile. Although La Palma is not located in a major tectonic earthquake zone, it is a volcanic island where seismic activity associated with volcanic processes is regularly observed.

Dynamic analyses performed using the Chilean seismic input predicted significant response accelerations at the camera and large fluctuations in the forces of the CSS tension ropes. Of particular concern was the possibility that some ropes could alternate between high tensile loads and complete loss of tension during a seismic event, potentially resulting in nonlinear structural behaviour and increased dynamic response. Mitigating these effects would have required modifications to the telescope structure.

Subsequently, a site-specific seismic hazard assessment for La Palma became available, defining a design response spectrum corresponding to a peak ground acceleration of approximately 0.05 g. Under these conditions, the predicted acceleration at the camera was approximately 0.6 g, a level considered fully acceptable for both the telescope structure and the supported instrumentation. Furthermore, all CSS ropes remained under tension throughout the analysed seismic events, eliminating concerns regarding rope slackening.

Consequently, no seismic modifications were required for the La Palma installation. However, the substantially higher seismic demands anticipated for the CTA South site in Chile later motivated the development of a modified LST design specifically optimized for severe seismic loading conditions. This design is described in a separate publication.

## CONCLUSIONS

This paper has presented the mechanical design, structural characteristics, and operational experience gained during the development and commissioning of the Large-Sized Telescope LST-1 at the Observatorio del Roque de los Muchachos on La Palma. The LST represents the largest and most dynamic telescope of the Cherenkov Telescope Array and was specifically designed to achieve the lowest energy threshold ever reached by a ground-based Cherenkov telescope while maintaining the rapid repositioning capability required for transient astrophysical events such as gamma-ray bursts.
The development of the telescope structure was carried out over an approximately eight-year design phase between 2008 and 2016. The resulting design combines an ultra-lightweight space-frame structure with high stiffness, excellent dynamic performance, and the robustness required to withstand the harsh environmental conditions encountered at the observatory, including extreme wind loads, ice accretion, and seismic events. The use of optimized steel, aluminium, and CFRP structural elements enabled the realization of a 23 m diameter telescope with a total moving mass of only approximately 110 tons while retaining the ability to rotate by 180° in less than 20 s.

LST-1 has been in operation since 2019 and has provided valuable validation of the design assumptions and analysis methods used during development. Operational experience has confirmed the overall suitability of the structural concept and has demonstrated that the telescope performs as intended. Several minor issues were identified during the early operational phase, including vortex-induced vibrations of selected structural members, the effects of severe icing events, and localized solar reflections from the segmented reflector. These issues were successfully mitigated through relatively simple engineering measures and did not require fundamental modifications to the structural concept.

The extensive analytical verification performed during the design phase, combined with several years of operational experience, has demonstrated that the telescope structure possesses adequate strength, stiffness, fatigue resistance, and fail-safe characteristics for long-term operation. The structural behaviour observed in service is in good agreement with the design predictions and confirms the validity of the adopted lightweight design philosophy.

The successful commissioning and operation of LST-1 provided the basis for the construction of three additional telescopes of essentially identical design. At the time of writing, LST-2, LST-3, and LST-4 are being assembled and commissioned. Together, the four telescopes will form the core of the CTA low-energy array and will constitute the most sensitive Cherenkov telescope system ever built in the energy range around 20 GeV. The unprecedented sensitivity of this four-telescope system will open a new observational window for the study of transient and distant gamma-ray sources and represents a major milestone in ground-based gamma-ray astronomy.

The successful realization of LST-1 demonstrates that extremely lightweight large-aperture telescope structures can achieve both outstanding scientific performance and long-term operational reliability, providing a solid foundation for future generations of high-performance Cherenkov telescopes.

## ACKNOWLEDGEMENTS

The authors would like to acknowledge the many individuals, institutions, and industrial partners whose contributions were essential to the successful development, construction, and commissioning of the Large-Sized Telescope (LST-1). The LST was designed, manufactured, assembled, and commissioned through the collaboration of the scientific institutes participating in the LST Consortium, supported by numerous industrial partners. The authors are particularly grateful to Eckard Lorenz (MPP) for his vision and guidance during the initial development of the telescope concept, and to Klaus Charne (MERO-TSK) for the outstanding collaboration in the design, engineering, manufacturing, and installation of the telescope structure.

We would also like to acknowledge the technical teams of the partner institutions for their highly constructive collaboration on subsystem integration and interface development:

- Universidad de Jaén, Spain: María Encarnación Garrido Ruiz.
- Istituto Nazionale di Fisica Nucleare (INFN), Italy: Adriano Pepato.

The authors further acknowledge the contributions of the engineering companies that supported the structural analyses and verification activities:

- AIRWORKS: Alessandro Targusi, Enrico Sambenedetto, Martina Cocciolo, and Luca Sillari.
- BURATTI: Enrico Buratti.
- GUIZZO SPACE: Gian Paolo Guizzo.

Special thanks are also due to the industrial partners who supplied major customized components and services for the telescope:

- MERO-TSK – primary telescope structure.
- MBM – azimuth rail system.
- FUTUR FIBRES – CFRP tether ropes of the Camera Support Structure.
- SCHORMAIR – access tower.
- BREVETTI STENDALTO – azimuth and elevation cable-chain systems.

The authors gratefully acknowledge the Max Planck Institute for Physics (MPP), Garching, Germany, and the Max Planck Society (Max-Planck-Gesellschaft) for their financial support of the structural design, analysis, development, and realization of the LST telescope. Their long-term commitment and support were instrumental in enabling the successful completion of this project.